\documentclass[runningheads]{llncs}
\usepackage[T1]{fontenc}
\usepackage{graphicx,verbatim}
\usepackage{amssymb}
\usepackage{multirow}
\usepackage{booktabs}
\usepackage{siunitx}
\usepackage[table]{xcolor}
\usepackage{threeparttable}
\begin{document}
\title{ICHOR: A Robust Representation Learning Approach for ASL CBF Maps with Self-Supervised Masked Autoencoders}
\titlerunning{ICHOR: Self-Supervised 3D MAE Pretraining for ASL CBF Maps}
%
\author{Xavier Beltran-Urbano\inst{1} \and
Yiran Li\inst{2} \and
Xinglin Zeng\inst{2} \and
Katie R. Jobson\inst{1} \and
Manuel Taso\inst{1} \and
Christopher A. Brown\inst{1} \and
David A. Wolk\inst{1} \and
Corey T. McMillan\inst{1} \and
Ilya M. Nashrallah\inst{1} \and
Paul A. Yushkevich\inst{1} \and
Ze Wang\inst{2} \and
John A. Detre\inst{1} \and
Sudipto Dolui\inst{1}
}
\authorrunning{Beltran-Urbano et al.}
%
\institute{University of Pennsylvania, Philadelphia, USA \and
University of Maryland School of Medicine, Baltimore, USA\\
\email{xurbano@seas.upenn.edu}}

  
\maketitle              
\begin{abstract}
Arterial spin labeling (ASL) perfusion MRI allows direct quantification of regional cerebral blood flow (CBF) without exogenous contrast, enabling noninvasive measurements that can be repeated without constraints imposed by contrast injection. ASL is increasingly acquired in research studies and clinical MRI protocols. Building on successes in structural imaging, recent efforts have implemented deep learning-based methods to improve image quality, enable automated quality control, and derive robust quantitative and predictive biomarkers with ASL-derived CBF. However, progress has been limited by variable image quality, substantial inter-site, vendor and protocol differences, and limited availability of labeled datasets needed to train models that generalize across cohorts. To address these challenges, we introduce ICHOR, a self-supervised pre-training approach for ASL CBF maps that learns transferable representations using 3D masked autoencoders. ICHOR is pre-trained via masked image modeling using a Vision Transformer backbone and can be used as a general-purpose encoder for downstream ASL tasks. For pre-training, we curated one of the largest ASL datasets to date, comprising 11,405 ASL CBF scans from 14 studies spanning multiple sites and acquisition protocols. We evaluated the pre-trained ICHOR encoder on three downstream diagnostic classification tasks and one ASL CBF map quality prediction regression task. Across all evaluations, ICHOR outperformed existing neuroimaging self-supervised pre-training methods adapted to ASL. Pre-trained weights and code will be made publicly available.


\end{abstract}

\section{Introduction}
Regional brain perfusion can be quantified via cerebral blood flow (CBF), defined as the volume of blood flowing through a specific region of brain tissue per unit time. CBF is a fundamental physiological marker of brain function that reflects cerebrovascular integrity and metabolic demand \cite{MonisPaper}, and is sensitive to both normal aging and a wide range of neurological disorders \cite{cbfchanges}. Conventional perfusion imaging typically relies on exogenous and often radioactive contrast agents, which can limit repeated use and may be contraindicated in vulnerable populations. In contrast, arterial spin labeling (ASL) perfusion MRI \cite{detre1992perfusion} provides noninvasive quantification of regional CBF using magnetically labeled arterial blood water as an endogenous tracer. ASL has demonstrated clinical utility in the evaluation and monitoring of Alzheimer’s disease (AD), cerebrovascular disorders, brain tumors, epilepsy, and other neurological conditions \cite{reviewASL}, where regional CBF provides a biomarker of pathology and disease progression \cite{biomarker}. ASL is also included in the imaging protocols of several large-scale neuroimaging studies, including ADNI \cite{ADNI}, UK Biobank \cite{UKbio}, and CLARiTI \cite{CLARiTI}. Despite this growing adoption, large-scale ASL analysis can be challenging due to intrinsically low signal and methodological variability \cite{reviewASLmethods}, which contribute to variable image quality and substantial inter-site and inter-protocol heterogeneity. This heterogeneity, together with the limited availability of labeled datasets, has hindered the development of deep learning (DL) methods in ASL imaging.

Self-supervised learning (SSL) offers a natural way to mitigate limited labeled data by pre-training models on large collections of unlabeled scans and then adapting them to specific downstream tasks with minimal annotation \cite{SSL-Medical}. In SSL, models learn transferable representations by solving surrogate objectives that exploit intrinsic structure in the data \cite{gui2024surveyselfsupervisedlearningalgorithms}. Among recent SSL strategies, masked autoencoders (MAEs) \cite{he2021maskedautoencodersscalablevision} have shown strong performance for volumetric imaging by randomly masking a large fraction of input patches and training the model to reconstruct the missing content from the visible context. This reconstruction objective encourages spatially coherent and semantically meaningful representations that can transfer well to tasks such as prediction, classification, and quality control. Vision Transformers (ViTs) are particularly well-suited to MAE-based pre-training, as their self-attention mechanism captures long-range dependencies and global context in 3D medical images. However, ViT-based models typically require large and diverse pre-training datasets to generalize well, an assumption that is often difficult to meet in medical imaging.

In neuroimaging, several large-scale pre-trained models have been proposed for brain MRI analysis, including BrainIAC \cite{BrainIAC} and  BrainSegFounder \cite{BRAINSEGFOUNDER}, and have demonstrated strong transfer to a range of downstream tasks. However, these efforts have largely focused on structural or multi-contrast anatomical MRI and have not incorporated physiological imaging modalities such as ASL. Consequently, ASL-derived CBF lacks a dedicated pre-trained backbone, despite its clinical relevance and increasing availability in large population studies. This gap motivates the development of pre-trained models tailored to ASL CBF maps that can support a broad range of downstream analyses.

\begin{figure} [t!]
\includegraphics[width=\textwidth]{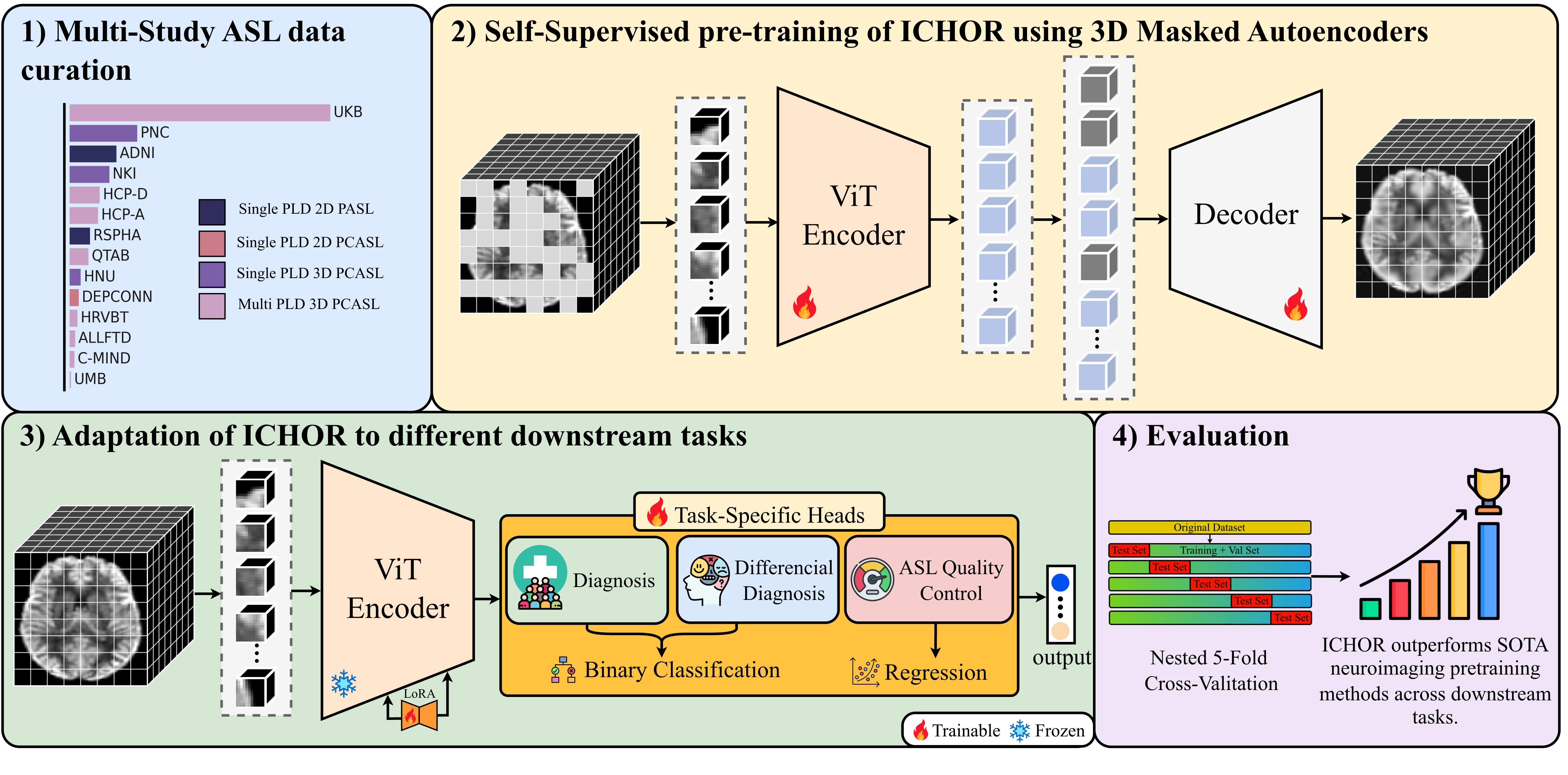}
\caption{Overview of the presented study.} \label{fig1}
\end{figure}

Here, we introduce ICHOR, a self-supervised pre-training framework for ASL CBF maps based on 3D MAEs. To support large-scale pre-training, we curated one of the largest ASL datasets to date, comprising 11,405 CBF scans from 14 studies spanning diverse sites and acquisition protocols. We evaluate the pre-trained ICHOR encoder on three diagnostic classification tasks and one ASL CBF map quality prediction task. Our contributions are: (1) a dedicated SSL pre-training framework for ASL CBF representation learning, (2) a large curated multi-site ASL dataset supporting pre-training across heterogeneous acquisition settings, and (3) an extensive downstream evaluation demonstrating improved performance over existing neuroimaging SSL baselines when adapted to ASL. An overview of this study is illustrated in Fig.~\ref{fig1}.
\section{Proposed Approach}

\begin{figure} [t!]
\centering
\includegraphics[width=\textwidth]{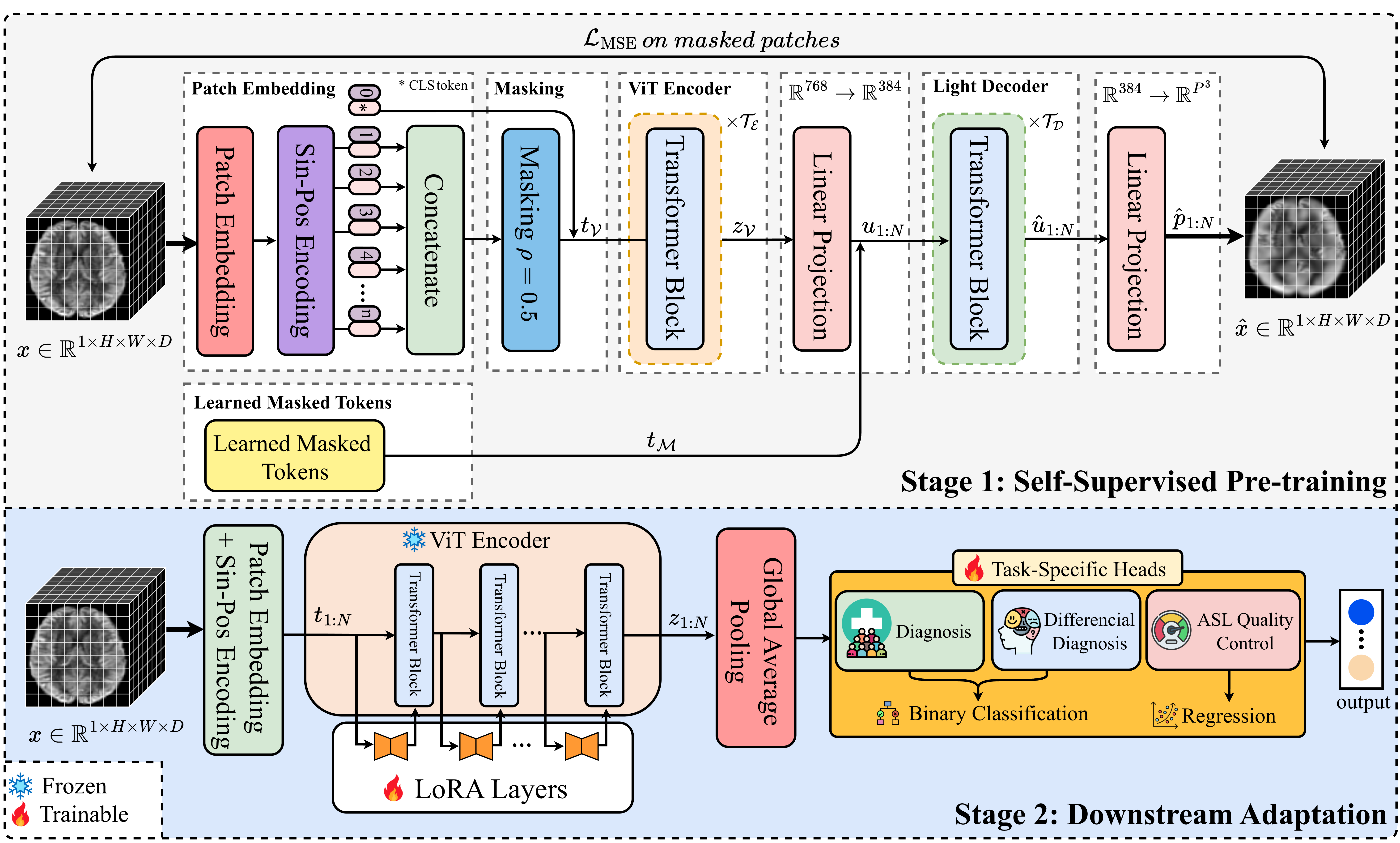}
\caption{Schematic of the ICHOR framework. Stage 1 (top) represents the proposed self-supervised 3D MAE pre-training pipeline. Stage 2 (bottom) illustrates the downstream task adaptation.} \label{fig2}
\end{figure}

Our framework adopts the pre-train-and-finetune paradigm for SSL approaches. First, we pre-train an asymmetric encoder-decoder architecture in a self-supervised manner by reconstructing randomly masked patches from the visible context. Next, we adapt the pre-trained encoder for downstream tasks using supervised learning. These two stages are described in the following sections.

\subsection{Stage 1: Self-Supervised Model Pre-Training}

As illustrated in Fig.~\ref{fig2}, the core objective of this stage is to learn robust representations by reconstructing randomly masked portions of the input. This stage can be divided into the following sections: 

\textbf{Model Input.} Let $x \in \mathbb{R}^{1 \times H \times W \times D}$ denote a preprocessed ASL CBF volume registered to the Montreal Neurological Institute (MNI) space and resampled to a fixed size of $(H,W,D)=(96,96,96)$. We split $x$ into $N=(H/P)(W/P)(D/P)=512$ non-overlapping 3D patches of size $P\times P\times P$ with $P=12$. We then randomly mask a fraction $\rho=0.5$ of the patches by selecting masked indices $\mathcal{M}\subset\{1,\dots,N\}$ with $|\mathcal{M}|=\rho N$, and denote the visible patches (unmasked patches) by $\mathcal{V}=\{1,\dots,N\}\setminus\mathcal{M}$.

\textbf{Vision Encoder.} The visible patches are embedded into tokens $t_\mathcal{V} \in \mathbb{R}^{|\mathcal{V}| \times 768}$ through a patch embedding layer. Then, sinusoidal positional embeddings are added to encode spatial information.  The resulting tokens are processed through a ViT-Base architecture comprising 12 transformer blocks ($T_{\mathcal{E}}$) with 12 attention heads and an MLP dimension of 3072, producing latent representations $z_\mathcal{V} \in \mathbb{R}^{|\mathcal{V}| \times 768}$.

\textbf{Light Decoder.} To reconstruct the masked content, we first project the encoder outputs to the decoder dimensions ($\mathbb{R}^{768} \rightarrow \mathbb{R}^{384}$). We then construct the full decoder input sequence $u_{1:N} \in \mathbb{R}^{N \times 384}$ by placing the projected visible tokens at their original positions and inserting learnable mask tokens at the masked positions $\mathcal{M}$. The resulting sequence is processed through a light-weight decoder comprising 4 transformer blocks ($T_{\mathcal{D}}$) with 12 attention heads and an MLP dimension of 1536, producing decoder representations $\hat{u}_{1:N} \in \mathbb{R}^{N \times 384}$. Finally, a linear projection layer ($\mathbb{R}^{384} \rightarrow \mathbb{R}^{P^3}$) maps the decoder tokens to predict the corresponding patch content $\hat{p}_{1:N} \in \mathbb{R}^{N \times P^3}$.

\textbf{Model optimization.} The model is optimized by minimizing a mean squared error (MSE) loss function computed only over the masked patches:
\begin{equation}
\mathcal{L}_{\text{MSE}} = \frac{1}{|\mathcal{M}|} \sum_{i \in \mathcal{M}} \left\| \hat{p}_i - p_i \right\|_2^2
\end{equation}
where $\hat{p}_i \in \mathbb{R}^{P^3}$ denotes the predicted patch and $p_i \in \mathbb{R}^{P^3}$ denotes the ground-truth patch.

\subsection{Stage 2: Downstream Adaptation}
After pre-training, we adapt the encoder to downstream tasks using Low-Rank Adaptation (LoRA) \cite{hu2021loralowrankadaptationlarge}, which enables efficient fine-tuning while mitigating catastrophic forgetting. As illustrated in Fig.~\ref{fig2}, trainable LoRA layers are inserted into the pre-trained ViT encoder while the original weights remain fixed. 

During downstream adaptation, no masking is applied, resulting in $N=512$ visible tokens. The pre-trained encoder processes the input tokens $t_\mathcal{V} \in \mathbb{R}^{|\mathcal{V}| \times 768}$ to produce latent representations $z_{1:N} \in \mathbb{R}^{N \times 768}$. Then, these representations are aggregated via global average pooling across spatial dimensions, and the resulting fixed-size feature vector is fed to a task-specific head consisting of one normalization layer followed by a fully connected layer. 

\section{Experimental Setup}
\subsection{Pre-training Configuration}
To train our model, we created a large multi-site ASL CBF map collection by aggregating data from publicly available and institutional studies (Table~\ref{tab:demographics}). In total, after undergoing quality control using an automated tool \cite{qei}, our dataset included 11{,}405 scans from 14 studies spanning diverse acquisition protocols and heterogeneous populations. Before model training, all scans were preprocessed following well-established standardized pipelines \cite{ASLtlbx,DOLUI2019101897} consisting of: (i) computation of CBF maps from raw data, (ii) spatial normalization to MNI space at $2\,mm^3$ resolution, (iii) cropping to a fixed brain bounding box to reduce background, and (iv) resampling to a fixed grid of $96\times 96\times96$. Finally, (v) CBF intensities were normalized by clipping to $[0,100]$ and dividing by 100, resulting in values within $[0,1]$. 

For model training, an NVIDIA GeForce RTX 4090 with 24GB memory was used for 400 epochs using AdamW optimizer with a base learning rate of $1.5\times10^{-4}$, weight decay of 0.05, and batch size of 48. We also applied cosine learning rate decay with 40 epochs of linear warmup. To mitigate dataset imbalance, we used weighted random sampling with soft balancing ($\alpha_{\text{bal}}=0.5$).

\subsection{Downstream Tasks and Evaluation}
For all downstream task adaptations, we inserted LoRA modules into the query, key, value, and output projection layers of each transformer block, using rank $r=8$, scaling factor $\alpha_{\text{LoRA}}=16$, and dropout $p=0.2$. Models were trained on an NVIDIA A100 40\,GB GPU for 100 epochs with a batch size of 8, using the AdamW optimizer with a base learning rate of $5\times10^{-4}$, weight decay of 0.05, and cosine learning rate decay with 10 epochs of linear warmup. A weighted binary cross-entropy (BCE) loss was used for classification tasks, with class weights inversely proportional to class frequencies, and MSE loss for the regression task.

Using this experimental setup, we evaluate ICHOR against state-of-the-art (SOTA) baselines, BrainIAC~\cite{BrainIAC}, BrainSegFounder~\cite{BRAINSEGFOUNDER}, and MedicalNet~\cite{chen2019med3dtransferlearning3d}, each pre-trained with structural MRI scans. Since MedicalNet is a ResNet-based architecture rather than a transformer, it was fine-tuned by updating all parameters rather than via LoRA modules, with all other hyperparameters kept identical across models. We evaluate across four downstream tasks: (1) binary classification of cognitively unimpaired amyloid-negative (CU A$\beta-$; $n=103$) versus cognitively impaired amyloid-positive participants (CI A$\beta+$; $n=44$), where amyloid positivity reflects Alzheimer's disease neuropathology; (2) ASL quality prediction, a regression task targeting a continuous score in $[0,1]$ ($n=383$), with a higher score reflecting a better quality CBF map; (3) binary classification of healthy older adults (HOA; $n=20$) versus small vessel disease patients (SVD; $n=15$); and (4) differential diagnosis between AD ($n=27$) and behavioral variant frontotemporal dementia (bvFTD; $n=36$).
\begin{figure}[t!]
\centering
\includegraphics[width=0.9\textwidth]{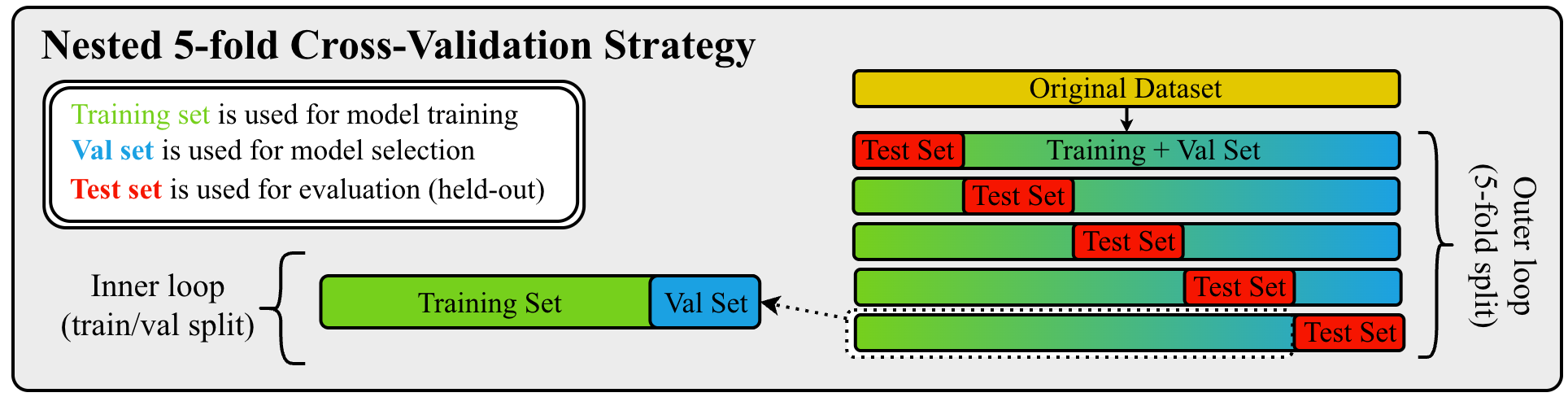}
\caption{Nested 5-fold cross-validation strategy used for evaluation.}
\label{fig3}
\end{figure}

Due to the limited sample size of some datasets, we employed a nested 5-fold cross-validation strategy to evaluate the models, as illustrated in Fig.~\ref{fig3}. Each dataset was randomly partitioned into five stratified folds to preserve class distribution across splits, and in each iteration, one fold was held out as an independent test set while the remaining folds were used for training, with an inner 80\%/20\% stratified train/validation split for model training and selection. Performance is reported as the mean across the five held-out test folds, using the area under the curve (AUC), accuracy, precision, recall, and F1-score for classification tasks, and MSE, root MSE (RMSE), mean absolute error (MAE), $R^2$ and Pearson correlation (PC) for regression tasks. All datasets used for downstream adaptation were obtained from institutional cohorts independent of the pre-training data, with demographic and acquisition details provided in Tables~\ref{tab:diagnosis_demographics_cu_ci}--\ref{tab:study_demographics}.

\begin{table}[t]
\centering
\caption{Results obtained by adapting the pre-trained encoder to downstream tasks using LoRA. Rand denotes randomly initialized model parameters. Arrows indicate desired direction. Metrics are reported as mean across the five outer folds. (Unit: \%)}
\label{tab:downstream_results}
\footnotesize
\setlength{\tabcolsep}{3.4pt}
\renewcommand{\arraystretch}{1.12}
\resizebox{\textwidth}{!}{%
\begin{tabular}{l|ccccc|ccccc}
\specialrule{2pt}{0pt}{0pt}
\multirow{2}{*}{Method}
& \multicolumn{5}{c|}{CU A$\beta-$  vs CI A$\beta+$}
& \multicolumn{5}{c}{ASL-Quality Prediction} \\
\cmidrule(lr){2-6} \cmidrule(lr){7-11}
& AUC $\uparrow$ & Accuracy $\uparrow$ & Precision $\uparrow$ & Recall $\uparrow$ & F1-score $\uparrow$ &
MSE $\downarrow$ &  MAE $\downarrow$ & RMSE $\downarrow$ & $R^2$ $\uparrow$ & PC $\uparrow$  
\\
\midrule
Rand                      & 68.24 & 57.12 & 38.5  & \textbf{74.44} & 50.34              & 1.21 & 8.14 & 10.96 & 83.99 & 92.03  \\
MedicalNet \cite{chen2019med3dtransferlearning3d}                & 70.31 & 68.80 & 51.07 & 56.94 &  51.80            & 1.31 & 8.29 & 11.32 & 82.54 & 91.21 \\
BrainSegFounder \cite{BRAINSEGFOUNDER}            & 65.90 & 59.90 & 40.54 & 72.5 &  51.28            & 1.21 & 8.25 & 10.94 & 83.86 & 91.98 \\
BrainIAC \cite{BrainIAC}                 & 55.97 & 57.93 & 37.86 & 54.72 & 41.71 & 1.44 & 8.72 & 11.97 & 80.94 & 90.30 \\
\textbf{Ours}             & \textbf{78.93} & \textbf{73.54} & \textbf{57.89} & 63.61 & \textbf{59.37} & \textbf{1.04} & \textbf{7.64} & \textbf{10.15} & \textbf{86.29} & \textbf{93.11} \\
\specialrule{2pt}{0pt}{0pt}
\multirow{2}{*}{Method}
& \multicolumn{5}{c|}{HOA vs SVD}
& \multicolumn{5}{c}{AD vs bvFTD} \\
\cmidrule(lr){2-6} \cmidrule(lr){7-11}
& AUC $\uparrow$ & Accuracy $\uparrow$ & Precision $\uparrow$ & Recall $\uparrow$ & F1-score $\uparrow$
& AUC $\uparrow$ & Accuracy $\uparrow$ & Precision $\uparrow$ & Recall $\uparrow$ & F1-score $\uparrow$ \\
\midrule
Rand                      & 55.83 & 42.85 & 40.95 & \textbf{86.67} & \textbf{54.46} & 84.52 & 81.28 & 86.17 & 83.93 & 83.99 \\
MedicalNet \cite{chen2019med3dtransferlearning3d}               & 48.33 & 48.57 & 38.57 & 53.33 &  41.33 &  98.02  & 89.10 & 94.16 & 86.78 & 89.61 \\
BrainSegFounder \cite{BRAINSEGFOUNDER}          & 48.33 & 51.43 & 28.57 & 40.00 & 31.43 & 85.80 & 82.94 & 85.27 & 86.67 & 85.45  \\
BrainIAC \cite{BrainIAC}                 & 50.00 & 54.29 & 41.91 & 60.00 & 48.19 & 86.11 &85.89& 93.05 & 83.92 & 86.67 \\
\textbf{Ours}             & \textbf{73.33} & \textbf{71.43} & \textbf{66.67} & 46.67 & 52.67 & \textbf{100.00}  &\textbf{ 98.33} & \textbf{100.00} & \textbf{97.14} & \textbf{98.46} \\
\specialrule{2pt}{0pt}{0pt}
\end{tabular}%
}
\end{table}

\section{Results and Discussion}
\subsection{Comparison with State-of-the-Art}
Results across all downstream tasks are presented in Table~\ref{tab:downstream_results}. Compared to the pre-trained baselines trained on structural imaging, ICHOR improves overall performance across all four tasks, with the strongest gains on the diagnostic classification problems. A likely contributor to these gains is modality mismatch during pre-training. Structural MRI encodes anatomical contrast, whereas ASL CBF maps capture a physiological perfusion signal. As a result, representations learned from structural MRI, including features related to tissue boundaries, morphological patterns, and intensity gradients, may transfer less effectively to perfusion-derived CBF distributions, which can be particularly limiting for diagnostic tasks with subtle inter-class differences (for example, HOA vs SVD).

This effect is less pronounced for the ASL quality prediction task, where structural MRI pre-trained baselines perform more competitively. Scan quality is influenced by low-level signal properties such as signal-to-noise ratio, smoothness, and intensity contrast, which are more transferable across MRI modalities than the disease-specific perfusion patterns required for diagnostic classification.

Finally, the randomly initialized baseline underperforms ICHOR on most metrics across the four tasks, though in some settings achieving higher recall and F1,  but performs competitively with the structural MRI pre-trained baselines. One likely reason is that the pre-trained models are adapted to ASL using LoRA, which may limit adaptation under strong modality mismatch, whereas the randomly initialized model is trained entirely on ASL data. Additionally, several downstream cohorts are small, and training from random initialization is prone to overfitting, which can inflate recall and F1. Metrics such as AUC, which are robust to such effects, confirm that ICHOR consistently outperforms random initialization, underscoring the value of modality-specific pre-training.

\subsection{Effect of Masking Ratio}

\begin{figure}[t]
\centering
\includegraphics[width=\textwidth]{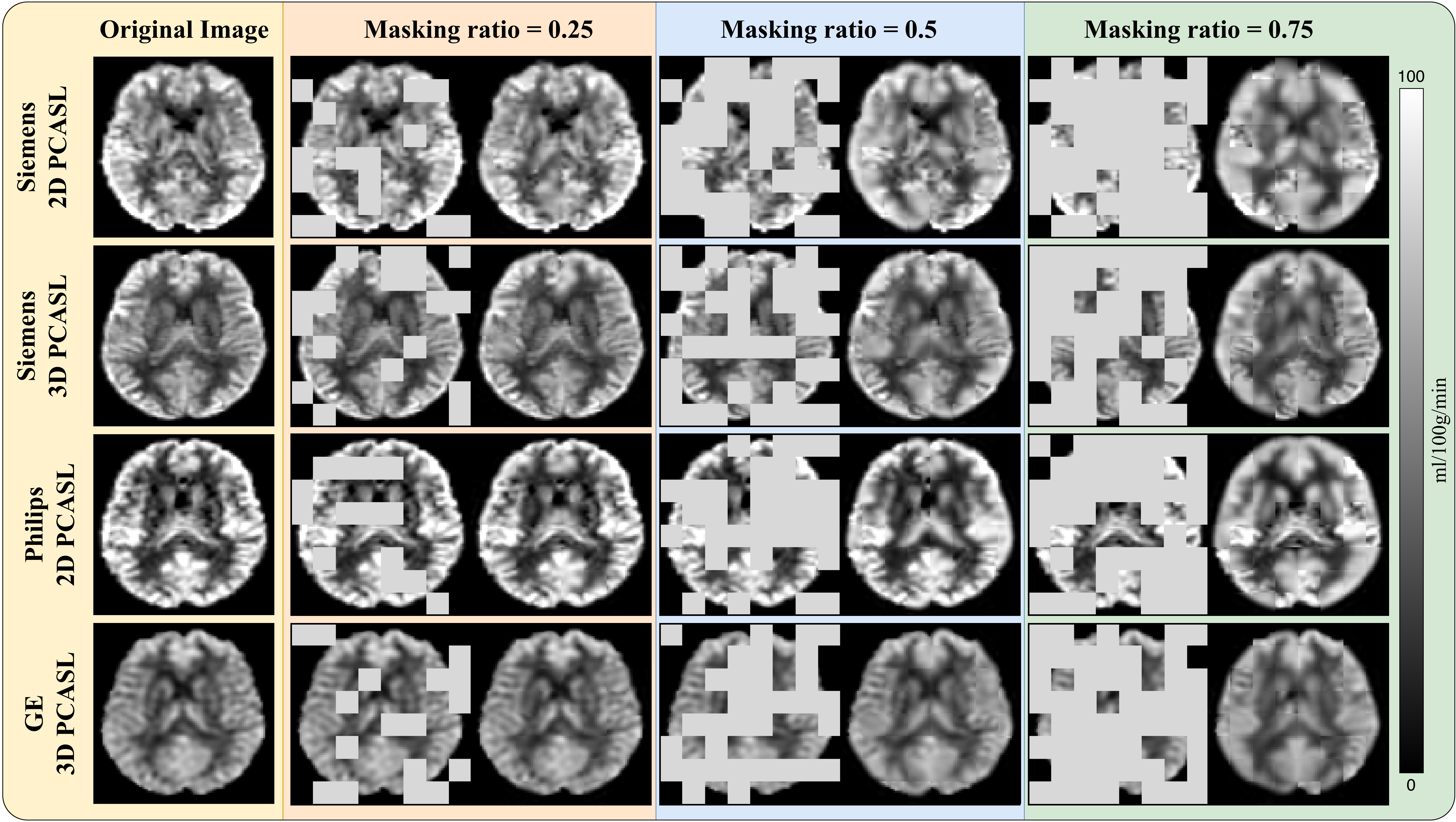}
\caption{Reconstruction results for pre-trained models trained with different masking ratios across imaging protocols and scanner vendors.}
\label{figrecons}
\end{figure}

We evaluated the effect of the masking ratio $\rho$ on both reconstruction quality and downstream task performance. As shown in Fig.~\ref{figrecons}, increasing $\rho$ yields visually sharper reconstructions across acquisition protocols and scanner vendors. However, downstream performance does not monotonically improve with reconstruction quality. To quantify this effect, we pre-trained ICHOR using $\rho \in \{0.25, 0.5, 0.75\}$ and adapted each pre-trained encoder to the CU A$\beta-$ vs CI A$\beta+$ classification task using the same downstream training protocol. As reported in Table~\ref{tab:rho_metrics}, $\rho=0.5$ achieves the best overall downstream performance across most metrics. This suggests a tradeoff in which low masking ratios make the reconstruction objective less informative, whereas high masking ratios provide insufficient visible context and make reconstruction overly challenging.

\begin{table}[t!]
\centering
\caption{Performance on CU A$\beta-$  vs CI A$\beta+$ task for different $\rho$ values. (Unit: \%)}
\label{tab:rho_metrics}
\scriptsize
\setlength{\tabcolsep}{6pt}
\renewcommand{\arraystretch}{1.15}
\begin{tabular}{lcccccc}
\specialrule{2pt}{0pt}{0pt}
\textbf{Metrics} & \textbf{AUC $\uparrow$} & \textbf{Accuracy $\uparrow$} & \textbf{Precision $\uparrow$} & \textbf{Recall $\uparrow$} & \textbf{F1-score $\uparrow$} \\
\midrule
$\rho=0.25$ & 76.74 & 69.54 & 55.42 & \textbf{65.55}  & 56.94  \\
$\rho=0.5$  & \textbf{78.93} & \textbf{73.54} & \textbf{57.90}& 63.61 & \textbf{59.37}  \\
$\rho=0.75$ & 71.90 & 69.44 & 50.32 & 61.66 & 54.67  \\
\specialrule{2pt}{0pt}{0pt}
\end{tabular}
\end{table}

\section{Conclusion}
We presented ICHOR, a self-supervised pre-training framework for ASL-derived CBF maps. Across four downstream tasks, ICHOR achieves the best overall performance relative to pre-trained neuroimaging baselines trained on structural MRI and adapted to ASL. By leveraging a large and diverse multi-study, multi-site ASL dataset for pre-training, ICHOR addresses the limited availability of labeled ASL data and provides a transferable encoder for downstream ASL analyses. Future directions include extending ICHOR to quality enhancement tasks such as denoising and artifact correction, leveraging ICHOR representations for longitudinal disease progression and treatment response modeling, and expanding the pre-training corpus to further strengthen generalization across diverse populations, protocols, and scanner platforms. 

\clearpage
\newpage

\bibliographystyle{splncs04}
\bibliography{bibliography}

@article{DOLUI2019101897,
title = {Characterizing a perfusion-based periventricular small vessel region of interest},
journal = {NeuroImage: Clinical},
volume = {23},
pages = {101897},
year = {2019},
issn = {2213-1582},
doi = {https://doi.org/10.1016/j.nicl.2019.101897},
url = {https://www.sciencedirect.com/science/article/pii/S2213158219302475},
author = {Sudipto Dolui and Dylan Tisdall and Marta Vidorreta and David R. Jacobs and Ilya M. Nasrallah and R. Nick Bryan and David A. Wolk and John A. Detre}
}

@article{MonisPaper,
author = {Taghvaei, Mohammad and Dolui, Sudipto and Sadaghiani, Shokufeh and Shakibajahromi, Banafsheh and Brown, Christopher and Khandelwal, Pulkit and Xie, Sharon X. and Das, Sandhitsu and Yushkevich, Paul A. and Wolk, David A. and Detre, John A.},
title = {Regional cerebral blood flow reflects both neurodegeneration and microvascular integrity across the Alzheimer's continuum},
journal = {Alzheimer's \& Dementia},
volume = {21},
number = {1},
pages = {e14382},
doi = {https://doi.org/10.1002/alz.14382},
url = {https://alz-journals.onlinelibrary.wiley.com/doi/abs/10.1002/alz.14382},
eprint = {https://alz-journals.onlinelibrary.wiley.com/doi/pdf/10.1002/alz.14382},
year = {2025}
}

@article{cbfchanges,
author = {Naghmeh Mokhber and Aidin Shariatzadeh and Abolfazl Avan and Hamidreza Saber and Golnaz Shojaeian Babaei and Gary Chaimowitz and M. Reza Azarpazhooh},
title ={Cerebral blood flow changes during aging process and in cognitive disorders: A review},

journal = {The Neuroradiology Journal},
volume = {34},
number = {4},
pages = {300-307},
year = {2021},
doi = {10.1177/19714009211002778},
    note ={PMID: 33749402},

URL = { 
    
        https://doi.org/10.1177/19714009211002778
    
    

},
eprint = { 
    
        https://doi.org/10.1177/19714009211002778
    
    

}
}

@article{reviewASLmethods,
author = {Hernandez-Garcia, Luis and Aramendía-Vidaurreta, Verónica and Bolar, Divya S. and Dai, Weiying and Fernández-Seara, Maria A. and Guo, Jia and Madhuranthakam, Ananth J. and Mutsaerts, Henk and Petr, Jan and Qin, Qin and Schollenberger, Jonas and Suzuki, Yuriko and Taso, Manuel and Thomas, David L. and van Osch, Matthias J. P. and Woods, Joseph and Zhao, Moss Y. and Yan, Lirong and Wang, Ze and Zhao, Li and Okell, Thomas W.},
title = {Recent Technical Developments in ASL: A Review of the State of the Art},
journal = {Magnetic Resonance in Medicine},
volume = {88},
number = {5},
pages = {2021-2042},
doi = {https://doi.org/10.1002/mrm.29381},
url = {https://onlinelibrary.wiley.com/doi/abs/10.1002/mrm.29381},
eprint = {https://onlinelibrary.wiley.com/doi/pdf/10.1002/mrm.29381},
year = {2022}
}

@article{CLARiTI,
author = {Mormino, Elizabeth C. and Biber, Sarah A. and Rahman-Filipiak, Annalise and Arfanakis, Konstantinos and Clark, Lindsay and Dage, Jeffrey L. and Detre, John A. and Dickerson, Bradford C. and Donohue, Michael C. and Kecskemeti, Steven and Hohman, Timothy J. and Jagust, William J. and Keene, Dirk C. and Kukull, Walter and Levendovszky, Swati R. and Rosen, Howie and Thompson, Paul M. and Villemagne, Victor L. and Wolk, David A. and Okonkwo, Ozioma C. and Rabinvovici, Gil D. and Rivera-Mindt, Monica and Foroud, Tatiana and Johnson, Sterling C.},
title = {The Consortium for Clarity in ADRD Research Through Imaging (CLARiTI)},
journal = {Alzheimer's \& Dementia},
volume = {21},
number = {1},
pages = {e14383},
doi = {https://doi.org/10.1002/alz.14383},
url = {https://alz-journals.onlinelibrary.wiley.com/doi/abs/10.1002/alz.14383},
eprint = {https://alz-journals.onlinelibrary.wiley.com/doi/pdf/10.1002/alz.14383},
year = {2025}
}

@article{SPRINT,
    author = {Dolui, Sudipto and Detre, John A. and Gaussoin, Sarah A. and Herrick, Jennifer S. and Wang, Danny J. J. and Tamura, Manjula Kurella and Cho, Monique E. and Haley, William E. and Launer, Lenore J. and Punzi, Henry A. and Rastogi, Anjay and Still, Carolyn H. and Weiner, Daniel E. and Wright, Jackson T., Jr and Williamson, Jeff D. and Wright, Clinton B. and Bryan, R. Nick and Bress, Adam P. and Pajewski, Nicholas M. and Nasrallah, Ilya M.},
    title = {Association of Intensive vs Standard Blood Pressure Control With Cerebral Blood Flow: Secondary Analysis of the SPRINT MIND Randomized Clinical Trial},
    journal = {JAMA Neurology},
    volume = {79},
    number = {4},
    pages = {380-389},
    year = {2022},
    month = {04},
    issn = {2168-6149},
    doi = {10.1001/jamaneurol.2022.0074},
    url = {https://doi.org/10.1001/jamaneurol.2022.0074},
    eprint = {https://jamanetwork.com/journals/jamaneurology/articlepdf/2789504/jamaneurology_dolui_2022_oi_220004_1649435699.28458.pdf},
}

@article{UKbio,
    doi = {10.1371/journal.pmed.1001779},
    author = {Sudlow, Cathie AND Gallacher, John AND Allen, Naomi AND Beral, Valerie AND Burton, Paul AND Danesh, John AND Downey, Paul AND Elliott, Paul AND Green, Jane AND Landray, Martin AND Liu, Bette AND Matthews, Paul AND Ong, Giok AND Pell, Jill AND Silman, Alan AND Young, Alan AND Sprosen, Tim AND Peakman, Tim AND Collins, Rory},
    journal = {PLOS Medicine},
    publisher = {Public Library of Science},
    title = {UK Biobank: An Open Access Resource for Identifying the Causes of a Wide Range of Complex Diseases of Middle and Old Age},
    year = {2015},
    month = {03},
    volume = {12},
    url = {https://doi.org/10.1371/journal.pmed.1001779},
    pages = {1-10},
    number = {3},

}

@article{CARDIA,
author = {Sudipto Dolui and Ze Wang and Danny JJ Wang and Raghav Mattay and Mack Finkel and Mark Elliott and Lisa Desiderio and Ben Inglis and Bryon Mueller and Randall B Stafford and Lenore J Launer and David R JacobsJr and R Nick Bryan and John A Detre},
title ={Comparison of non-invasive MRI measurements of cerebral blood flow in a large multisite cohort},

journal = {Journal of Cerebral Blood Flow \& Metabolism},
volume = {36},
number = {7},
pages = {1244-1256},
year = {2016},
doi = {10.1177/0271678X16646124},
    note ={PMID: 27142868},

URL = { 
    
        https://doi.org/10.1177/0271678X16646124
},
eprint = { 
    
        https://doi.org/10.1177/0271678X16646124
}
}

@article{
MESA,
author = {Thomas R. Austin  and Ilya M. Nasrallah  and Guray Erus  and Lisa M. Desiderio  and Lin Y. Chen  and Philip Greenland  and Barbara N. Harding  and Timothy M. Hughes  and Paul N. Jensen  and WT Longstreth  and Wendy S. Post  and Steven J. Shea  and Colleen M. Sitlani  and Christos Davatzikos  and Mohamad Habes  and R. Nick Bryan  and Susan R. Heckbert },
title = {Association of Brain Volumes and White Matter Injury With Race, Ethnicity, and Cardiovascular Risk Factors: The Multi‐Ethnic Study of Atherosclerosis},
journal = {Journal of the American Heart Association},
volume = {11},
number = {7},
pages = {e023159},
year = {2022},
doi = {10.1161/JAHA.121.023159},
URL = {https://www.ahajournals.org/doi/abs/10.1161/JAHA.121.023159},
eprint = {https://www.ahajournals.org/doi/pdf/10.1161/JAHA.121.023159}}

@article{
ADNI,
author = {R. C. Petersen  and P. S. Aisen  and L. A. Beckett  and M. C. Donohue  and A. C. Gamst  and D. J. Harvey  and C. R. Jack  and W. J. Jagust  and L. M. Shaw  and A. W. Toga  and J. Q. Trojanowski  and M. W. Weiner },
title = {Alzheimer's Disease Neuroimaging Initiative (ADNI)},
journal = {Neurology},
volume = {74},
number = {3},
pages = {201-209},
year = {2010},
doi = {10.1212/WNL.0b013e3181cb3e25},
URL = {https://www.neurology.org/doi/abs/10.1212/WNL.0b013e3181cb3e25},
eprint = {https://www.neurology.org/doi/pdf/10.1212/WNL.0b013e3181cb3e25}}

@article{SSL-Medical,
author = {Zeng, Xiangrui and Abdullah, Nibras and Putra, Sumari},
year = {2024},
month = {10},
pages = {},
title = {Self-supervised learning framework application for medical image analysis: a review and summary},
volume = {23},
journal = {BioMedical Engineering OnLine},
doi = {10.1186/s12938-024-01299-9}
}

@misc{gui2024surveyselfsupervisedlearningalgorithms,
      title={A Survey on Self-supervised Learning: Algorithms, Applications, and Future Trends}, 
      author={Jie Gui and Tuo Chen and Jing Zhang and Qiong Cao and Zhenan Sun and Hao Luo and Dacheng Tao},
      year={2024},
      eprint={2301.05712},
      archivePrefix={arXiv},
      primaryClass={cs.LG},
      url={https://arxiv.org/abs/2301.05712}, 
}

@misc{he2021maskedautoencodersscalablevision,
      title={Masked Autoencoders Are Scalable Vision Learners}, 
      author={Kaiming He and Xinlei Chen and Saining Xie and Yanghao Li and Piotr Dollár and Ross Girshick},
      year={2021},
      eprint={2111.06377},
      archivePrefix={arXiv},
      primaryClass={cs.CV},
      url={https://arxiv.org/abs/2111.06377}, 
}

@misc{chen2019med3dtransferlearning3d,
      title={Med3D: Transfer Learning for 3D Medical Image Analysis}, 
      author={Sihong Chen and Kai Ma and Yefeng Zheng},
      year={2019},
      eprint={1904.00625},
      archivePrefix={arXiv},
      primaryClass={cs.CV},
      url={https://arxiv.org/abs/1904.00625}, 
}

@article {BrainIAC,
	author = {Tak, Divyanshu and Garomsa, Biniam A. and Chaunzwa, Tafadzwa L. and Zapaishchykova, Anna and Climent Pardo, Juan Carlos and Ye, Zezhong and Zielke, John and Ravipati, Yashwanth and Vajapeyam, Sri and Mahootiha, Maryam and Smith, Ceilidh and Familiar, Ariana M. and Liu, Kevin X. and Prabhu, Sanjay and Bandopadhayay, Pratiti and Nabavizadeh, Ali and Mueller, Sabine and Aerts, Hugo JWL and Huang, Raymond Y. and Poussaint, Tina Y. and Kann, Benjamin H.},
	title = {A foundation model for generalized brain MRI analysis},
	elocation-id = {2024.12.02.24317992},
	year = {2024},
	doi = {10.1101/2024.12.02.24317992},
	publisher = {Cold Spring Harbor Laboratory Press},
	URL = {https://www.medrxiv.org/content/early/2024/12/03/2024.12.02.24317992},
	eprint = {https://www.medrxiv.org/content/early/2024/12/03/2024.12.02.24317992.full.pdf},
	journal = {medRxiv}
}

@misc{BRAINSEGFOUNDER,
      title={BrainSegFounder: Towards 3D Foundation Models for Neuroimage Segmentation}, 
      author={Joseph Cox and Peng Liu and Skylar E. Stolte and Yunchao Yang and Kang Liu and Kyle B. See and Huiwen Ju and Ruogu Fang},
      year={2024},
      eprint={2406.10395},
      archivePrefix={arXiv},
      primaryClass={eess.IV},
      url={https://arxiv.org/abs/2406.10395}, 
}

@article{detre1992perfusion,
  title={Perfusion imaging},
  author={Detre, John A and Leigh, John S and Williams, Donald S and Koretsky, Alan P},
  journal={Magnetic resonance in medicine},
  volume={23},
  number={1},
  pages={37--45},
  year={1992},
  publisher={Wiley Online Library}
}

@article{biomarker,
  title={Arterial spin labeling MRI: an emerging biomarker for Alzheimer's disease and other neurodegenerative conditions},
  author={Wolk, David A and Detre, John A},
  journal={Current opinion in neurology},
  volume={25},
  number={4},
  pages={421--428},
  year={2012},
  publisher={LWW}
}

@article{epilepsy,
  title={Detection of mesial temporal lobe hypoperfusion in patients with temporal lobe epilepsy by use of arterial spin labeled perfusion MR imaging},
  author={Wolf, Ronald L and Alsop, David C and Levy-Reis, Igor and Meyer, Philipp T and Maldjian, Joseph A and Gonzalez-Atavales, Julio and French, Jaqueline A and Alavi, Abass and Detre, John A},
  journal={American journal of neuroradiology},
  volume={22},
  number={7},
  pages={1334--1341},
  year={2001},
  publisher={American Journal of Neuroradiology}
}

@article{reviewASL,
title = {Clinical Neuroimaging Using Arterial Spin-Labeled Perfusion Magnetic Resonance Imaging},
journal = {Neurotherapeutics},
volume = {4},
number = {3},
pages = {346-359},
year = {2007},
note = {Advances in Neuroimaging/Neuroethics},
issn = {1878-7479},
doi = {https://doi.org/10.1016/j.nurt.2007.04.005},
url = {https://www.sciencedirect.com/science/article/pii/S1878747923006360},
author = {Ronald L. Wolf and John A. Detre}
}

@article{ASLtlbx,
title = {Empirical optimization of ASL data analysis using an ASL data processing toolbox: ASLtbx},
journal = {Magnetic Resonance Imaging},
volume = {26},
number = {2},
pages = {261-269},
year = {2008},
issn = {0730-725X},
doi = {https://doi.org/10.1016/j.mri.2007.07.003},
url = {https://www.sciencedirect.com/science/article/pii/S0730725X07003517},
author = {Ze Wang and Geoffrey K. Aguirre and Hengyi Rao and Jiongjiong Wang and María A. Fernández-Seara and Anna R. Childress and John A. Detre}
}

@inproceedings{qei,
author = {Urbano, Xavier and Taso, Manuel and Nasrallah, Ilya and Detre, John and Wang, Ze and Dolui, Sudipto},
pages = {},
title = {QEI-Net: A Deep learning-based automated quality evaluation index for ASL CBF Maps},
doi = {10.58530/2025/0730}
}

@misc{hu2021loralowrankadaptationlarge,
      title={LoRA: Low-Rank Adaptation of Large Language Models}, 
      author={Edward J. Hu and Yelong Shen and Phillip Wallis and Zeyuan Allen-Zhu and Yuanzhi Li and Shean Wang and Lu Wang and Weizhu Chen},
      year={2021},
      eprint={2106.09685},
      archivePrefix={arXiv},
      primaryClass={cs.CL},
      url={https://arxiv.org/abs/2106.09685}, 
}
\clearpage
\appendix
\renewcommand{\thetable}{A\arabic{table}}
\setcounter{table}{0}

\section{Appendix}
\begin{table}
\centering
\caption{Demographic information of included studies for pre-training.}
\label{tab:demographics}
\footnotesize
\setlength{\tabcolsep}{2.5pt}
\renewcommand{\arraystretch}{1.15}
\rowcolors{3}{gray!6}{white}
\begin{tabular}{
l
S[table-format=5.0]
S[table-format=3.0]
S[table-format=3.1]
S[table-format=2.1]
S[table-format=2.1]
S[table-format=3.1]
l
l
l
}
\toprule
& & \multicolumn{3}{c}{Age (years)} & \multicolumn{2}{c}{Participants (\%)} & \multicolumn{3}{c}{Protocol} \\
\cmidrule(lr){3-5} \cmidrule(lr){6-7} \cmidrule(lr){8-10}
{Dataset} & {N} & {Range} & {Mean} & {SD} & {Female} & {Controls} & {PLD} & {Dim} & {Labeling} \\
\midrule
ADNI     &  966 & {55--94}  & 72.7 &  7.2 & 56.4 &  31.2 & Single & 2D & PASL \\
RSPHA      &  422 & {56--91}  & 72.5 &  7.6 & 59.5 &  55.5 & Single & 2D & PASL \\
ALLFTD   &  114 & {18--84}  & 52.7 & 15.3 & 49.1 &  48.2 & Multi & 3D & PCASL \\
C-MIND   &  100 & {0--19}   &  9.3 &  5.4 & 49.0 & 100.0 & Multi & 3D & PCASL \\
DEPCONN  &  187 & {20--74}  & 35.5 & 11.7 & 55.1 &   0.0 & Single & 2D & PCASL \\
HCP-A   &  585 & {36--100} & 58.3 & 14.8 & 60.3 & 100.0 & Multi & 3D & PCASL \\
HCP-D   &  623 & {6--22}   & 14.4 &  4.0 & 53.8 & 100.0 & Multi & 3D & PCASL \\
HNU      &  228 & {5--61}   & 25.4 &  8.7 & 57.9 & 100.0 & Single & 3D & PCASL \\
HRVBT    &  163 & {18--80}  & 37.7 & 20.8 & 56.4 & 100.0 & Multi & 3D & PCASL \\
NKI      &  820 & {6--85}   & 38.3 & 21.5 & 60.6 & 100.0 & Single & 3D & PCASL \\
PNC      & 1397 & {8--23}   & 14.9 &  3.7 & 53.3 & 100.0 & Single & 3D & PCASL \\
QTAB     &  390 & {8--14}   & 10.8 &  1.3 & 48.2 & 100.0 & Multi & 3D & PCASL \\
UKB      & 5383 & {46--83}  & 63.0 &  8.1 & 53.5 & 100.0 & Multi & 3D & PCASL \\
UMB      &   27 & {51--83}  & 66.3 & 10.9 & 51.9 &  40.7 & Multi & 3D & PCASL \\
\midrule
\rowcolor{gray!12}
All      & 11405 & {0--100} & 49.7 & 23.6 & 54.7 &  90.2 & Mixed & Mixed & Mixed \\
\bottomrule
\end{tabular}
\scriptsize
\begin{minipage}{0.98\linewidth}
\textit{Acronyms:}
\textbf{ADNI}: Alzheimer’s Disease Neuroimaging Initiative;
\textbf{RSPHA}: Resting State Perfusion in Healthy Aging;
\textbf{ALLFTD}: ARTFL LEFFTDS Longitudinal Frontotemporal Lobar Degeneration;
\textbf{C-MIND}: Cincinnati MR Imaging of Neurodevelopment;
\textbf{HCP-A}: Human Connectome Project -- Aging;
\textbf{HCP-D}: Human Connectome Project -- Development;
\textbf{NKI}: Nathan Kline Institute -- Rockland Sample;
\textbf{PNC}: Philadelphia Neurodevelopmental Cohort;
\textbf{QTAB}: Queensland Twin Adolescent Brain;
\textbf{UKB}: UK Biobank;
\textbf{DEPCONN}: Depression Connectome project;
\textbf{HNU}: Hangzhou Normal University local collection;
\textbf{HRVBT}: Heart rate variability biofeedback training and emotion regulation;
\textbf{UMB}: University of Maryland, Baltimore local collection.
\end{minipage} 
\end{table}

\begin{table}
\centering
\caption{Demographic and acquisition information for the participants included for discriminating CU A$\beta-$  vs CI A$\beta+$.}
\label{tab:diagnosis_demographics_cu_ci}
\footnotesize
\setlength{\tabcolsep}{2.5pt}
\renewcommand{\arraystretch}{1.15}
\rowcolors{3}{gray!6}{white}
\begin{tabular}{p{0.58\linewidth}ccc}
\toprule
& \multicolumn{3}{c}{Diagnosis} \\
\cmidrule(lr){2-4}
& {CU (n=103)} & {MCI (n=35)} & {AD (n=9)} \\
\midrule
\rowcolor{gray!12}\multicolumn{4}{l}{\textbf{Demographics and biomarkers}} \\
Age (mean $\pm$ SD) & $71.2 \pm 5.7$ & $72.5 \pm 6.4$ & $71.6 \pm 7.4$ \\
Female sex, n (\%) & 68 (66.0) & 19 (54.3) & 4 (55.6) \\
Amyloid status (A$\beta$+/A$\beta$--) & 0/103 & 35/0 & 9/0 \\
\addlinespace[2pt]
\rowcolor{gray!12}\multicolumn{4}{l}{\textbf{Acquisition characteristics}} \\
PLD (Single/Multi) & 103/0 & 35/0 & 9/0 \\
Dimensions (2D/3D) & 11/92 & 5/30 & 0/9 \\
Protocol (PASL/PCASL) & 0/103 & 0/35 & 0/9 \\
\bottomrule
\end{tabular}
\end{table}

\begin{table}
\centering
\caption{Demographic and acquisition information for the participants included for discriminating HOA vs SVD.}
\label{tab:diagnosis_demographics_hoa_svd}
\footnotesize
\setlength{\tabcolsep}{2.5pt}
\renewcommand{\arraystretch}{1.15}
\rowcolors{3}{gray!6}{white}
\begin{tabular}{p{0.58\linewidth}cc}
\toprule
& \multicolumn{2}{c}{Diagnosis} \\
\cmidrule(lr){2-3}
& {Healthy older adults (n=20)} & {SVD (n=15)} \\
\midrule
\rowcolor{gray!12}\multicolumn{3}{l}{\textbf{Demographics}} \\
Age (mean $\pm$ SD) & $72.3 \pm 5.1$ & $64.3 \pm 5.2$ \\
Female sex, n (\%) & 9 (45.0) & 9 (60.0) \\
\addlinespace[2pt]
\rowcolor{gray!12}\multicolumn{3}{l}{\textbf{Acquisition characteristics}} \\
PLD (Single/Multi) & 20/0 & 15/0 \\
Dimensions (2D/3D) & 0/20 & 0/15 \\
Protocol (PASL/PCASL) & 0/20 & 0/15 \\
\bottomrule
\end{tabular}
\end{table}

\begin{table}
\centering
\caption{Demographic and acquisition information for the participants included for discriminating AD vs bvFTD.}
\label{tab:diagnosis_demographics_ad_bvftd}
\footnotesize
\setlength{\tabcolsep}{2.5pt}
\renewcommand{\arraystretch}{1.15}
\rowcolors{3}{gray!6}{white}
\begin{tabular}{p{0.58\linewidth}cc}
\toprule
& \multicolumn{2}{c}{Diagnosis} \\
\cmidrule(lr){2-3}
& {AD (n=27)} & {bvFTD (n=36)} \\
\midrule
\rowcolor{gray!12}\multicolumn{3}{l}{\textbf{Demographics}} \\
Age (mean $\pm$ SD) & $71.0 \pm 7.6$ & $64.0 \pm 7.4$ \\
Female sex, n (\%) & 14 (51.9) & 12 (33.33) \\
\addlinespace[2pt]
\rowcolor{gray!12}\multicolumn{3}{l}{\textbf{Acquisition characteristics}} \\
PLD (Single/Multi) & 27/0 & 36/0 \\
Dimensions (2D/3D) & 0/27 & 0/36\\
Protocol (PASL/PCASL) & 0/27 & 0/36 \\
\bottomrule
\end{tabular}
\end{table}

\begin{table}[t!]
\centering
\caption{Demographic and acquisition information for the participants included for predicting ASL quality.}
\label{tab:study_demographics}
\scriptsize
\setlength{\tabcolsep}{3pt}
\renewcommand{\arraystretch}{1.15}
\begin{threeparttable}
\rowcolors{3}{gray!6}{white}
\begin{tabular}{p{2.2cm}lcclcc}
\toprule
\textbf{Study} & \textbf{Scanner} & \textbf{N} & \textbf{Age (mean $\pm$ SD)} & \textbf{Female, n (\%)} & \textbf{Protocol} & \textbf{PLD} \\
\midrule
\rowcolor{gray!12}\textbf{SPRINT} & & 111 & $68.7 \pm 9.0$ & 41 (37) & 2D/3D PCASL & single \\
 & GE & 30 & $74.2 \pm 7.7$ & 7 (23) & 3D PCASL & single \\
 & Philips & 50 & $67.3 \pm 8.9$ & 21 (42) & 2D PCASL & single \\
 & Siemens & 31 & $65.5 \pm 8.2$ & 13 (42) & 2D PCASL & single \\
\addlinespace[2pt]
\rowcolor{gray!12}\textbf{CARDIA} & & 19 & $50.5 \pm 3.5$ & 12 (63) & 2D PCASL & single \\
 & Siemens & 19 & $50.5 \pm 3.5$ & 12 (63) & 2D PCASL & single \\
\addlinespace[2pt]
\rowcolor{gray!12}\textbf{VCID} & & 85 & $57.6 \pm 10.6$ & 50 (59) & 3D PCASL & single \\
 & GE & 40 & $61.2 \pm 11.0$ & 24 (60) & 3D PCASL & single \\
 & Siemens & 45 & $54.4 \pm 9.2$ & 26 (58) & 3D PCASL & single \\
\addlinespace[2pt]
\rowcolor{gray!12}\textbf{NACC} & & 55 & $69.1 \pm 13.4$ & 33 (60) & 2D/3D PCASL & single \\
 & Siemens & 55 & $69.1 \pm 13.4$ & 33 (60) & 2D/3D PCASL & single \\
\addlinespace[2pt]
\rowcolor{gray!12}\textbf{DLBS} & & 25 & $54.9 \pm 22.6$ & 16 (64) & 2D PCASL & single \\
 & Philips & 25 & $54.9 \pm 22.6$ & 16 (64) & 2D PCASL & single \\
\addlinespace[2pt]
\rowcolor{gray!12}\textbf{MESA} & & 45 & $66.9 \pm 8.6$ & 25 (56) & 3D PCASL & single \\
 & Siemens & 45 & $66.9 \pm 8.6$ & 25 (56) & 3D PCASL & single \\
\addlinespace[2pt]
\rowcolor{gray!12}\textbf{CLARiTI} & & 10 & $79.0 \pm 4.1$ & 7 (70) & 3D PCASL & single \\
 & Siemens & 10 & $79.0 \pm 4.1$ & 7 (70) & 3D PCASL & single \\
\addlinespace[2pt]
\rowcolor{gray!12}\textbf{Institutional data} & & 33 & $44.2 \pm 18.5$ & 16 (48) & 3D PCASL & single \\
 & Siemens & 33 & $44.2 \pm 18.5$ & 16 (48) & 3D PCASL & single \\
\midrule
\textbf{Total} & & 383 & $62.4 \pm 14.7$ & 200 (52) & 2D/3D PCASL & single \\
\bottomrule
\end{tabular}
\begin{tablenotes}[flushleft]
\scriptsize
\item[] \textit{Acronyms:}
\textbf{SPRINT}: Systolic Blood Pressure Intervention Trial;
\textbf{CARDIA}: Coronary Artery Risk Development in Young Adults;
\textbf{VCID}: Vascular Contributions to Cognitive Impairment and Dementia;
\textbf{NACC}: National Alzheimer's Coordinating Center;
\textbf{DLBS}: Dallas Lifespan Brain Study;
\textbf{MESA}: Multi-Ethnic Study of Atherosclerosis;
\textbf{CLARiTI}: Consortium for Clarity in ADRD Research Through Imaging;
\textbf{GE}: General Electric;
\textbf{SD}: Standard Deviation;
\textbf{PCASL}: Pseudo-Continuous Arterial Spin Labeling;
\textbf{PLD}: Post-Labeling Delay.
\end{tablenotes}
\end{threeparttable}
\end{table}

\end{document}